\documentclass[12pt,a4paper]{article}
\usepackage{epsfig}
\usepackage{amsmath}
\usepackage{amsfonts}
\usepackage{amssymb}
\usepackage{graphicx}

\begin{document}

\begin{flushright}
{\small LPTh-Ji 10/004}
\end{flushright}

\begin{center}
\vspace{2cm}

{\Large \textbf{\textit{Electroweak Phase Transition in the U(1)'-MSSM}}}

\vspace{1.3cm}

\textbf{Amine Ahriche}${}^{1}$ and \textbf{Salah Nasri}${}^{2}$

\vspace{0.6cm}

\textit{${}^{1}$ Laboratory of Theoretical Physics, Department of Physics,
University of Jijel, PB 98 Ouled Aissa, DZ-18000 Jijel, Algeria. }

\textit{${}^{2}$ Department of Physics, UAE University, P.O. Box 17551,
Al-Ain, United Arab Emirates. }

\textit{Emails: ahriche@univ-jijel.dz, snasri@uaeu.ac.ae.}
\end{center}

\vspace{1.2cm}

\hrule \vspace{0.5cm} {\normalsize \textbf{{\large {Abstract}}}}

\vspace{0.5cm}

In this work, we have investigated the nature of the electroweak phase
transition in the $U(1)$ extended minimal supersymmetric standard model
without introducing any exotic fields. The effective potential has been
estimated exactly at finite temperature taking into account the whole
particle spectrum. For reasonable values of the lightest Higgs and
neutralino, we found that the electroweak phase transition could be strongly
first-order due to: (1) the interactions of the singlet with the doublets in
the effective potential, and (2) the evolution of the wrong vacuum, that
delays the transition.

\vspace{0.8cm} \textbf{Keywords}: baryogenesis, electroweak phase
transition, extra singlet, extra gauge boson. \vspace{0.5cm}

\textbf{PACS}: 12.60.-i, 12.60.Jv, 14.80.Cp, 12.15.-y. \vspace{0.5cm} \hrule%
\vspace{2cm}

\section{Introduction}

The matter-antimatter asymmetry in the Universe is observed to be $%
n_{b}/n_{\gamma }\sim 10^{-10}$ \cite{wmap}. If this asymmetry is to be
explained by microphysics rather than initial conditions, then there must be
processes occurring in the early Universe that violate baryon number and CP
and which occur out of thermal equilibrium. It appears that the standard
model (SM) satisfies all three conditions \cite{sak, EB}; the baryon number
is not conserved at quantum level due to the $B+L$ anomaly \cite{B+L}, a CP
violation source does exist in the quark sector ($CKM$ matrix), and a
departure from thermal equilibrium could in principle be achieved through a
strong first-order phase transition \cite{EB}. However, detail calculations
show that the SM fails to generate the observed baryon asymmetry due to the
smallness of the $CP$ violation effect and the weakness of the electroweak
phase transition (EWPT) \cite{RT}.

In gauge theories, a first-order phase transition takes place if the vacuum
of the theory does not correspond to the global minimum of the potential.
Since it is energetically unfavored, the field changes its value to the true
vacuum ( i.e., the absolute minimum of the potential). Because of the
existence of a barrier between the two minima, this mechanism can happen by
tunneling or thermal fluctuations via bubble nucleation. The electroweak
baryogenesis scenario is realized when the $B$ and $CP$ violating
interactions pass through the bubble wall. These interactions are very fast
outside the bubbles but suppressed inside. Then a net baryon asymmetry
results inside the bubbles which are expanding and filling the Universe at
the end.

In the SM, the EWPT is too weak \cite{andhall} unless the Higgs mass is less
than $45$ GeV \cite{mhbound}, which is in conflict with present data \cite%
{LEP}. But a departure from thermal equilibrium without being in conflict
with this severe bound on the Higgs mass, is possible when extending the SM
with additional gauge singlets \cite{sms,amin}, new heavy fermions \cite{CC}%
, an extra Higgs doublet \cite{2H}, or in some supersymmetric extensions of
the SM.

In spite of its success and popularity, the minimal supersymmetric standard
model (MSSM) with R-parity still has two major problems: the $\mu $-problem
\cite{mu} and the fast proton decay due to dimension 5 operators \cite%
{pdecay}. A natural solution to these problems would probably require the
extension of MSSM by a new mechanism or a new symmetry. The $U(1)^{\prime }$%
-extended MSSM ($USSM$, $UNMSSM$\ or $UMSSM$) \cite{UMSSM} is a
straightforward extension of the MSSM with a nonanomalous TeV scale Abelian
gauge symmetry. This simple enlargement of the gauge sector is well
motivated in string construction \cite{USt}, in grand unified theories \cite%
{UGUT} such as $SO(10)$ and $E_{6}$, in models of dynamical symmetry
breaking \cite{USD} and little Higgs models \cite{ULT}. The $\mu $-problem
and the dangerous dimension 5 operator can be solved naturally with an
appropriate $U(1^{\prime })$ charge assignment. Furthermore, these $%
U(1)^{\prime }$ models can provide a new candidates for dark matter that are
not excluded by direct dark matter searches and with interesting signatures
at colliders \cite{UDM,UDM2}.

In the MSSM, the EWPT could be strongly first-order if the light stop is
lighter than the top quark \cite{MSSMPT}. In the singlet extended MSSM \cite%
{LLee}, such as the NMSSM \cite{NMSSM}, the EWPT get stronger easily for a
large range of parameters \cite{NMSSMPT}. In gauge extensions of the MSSM,
such as UMSSM, the EWPT is also strongly first-order but with the price of
introducing 3 new extra singlet scalars \cite{lang}, or by adding new extra
heavy singlet fermions \cite{ham}.

The main reason that makes it less easier to have a strong EWPT in the UMSSM
compared to NMSSM , is that the former contains a new gauge interaction
which results in strong constraint on the mixing between the SM gauge boson $%
Z$ and the new one $Z^{\prime}$ \cite{MMz}
\begin{equation}
2M_{ZZ^{\prime}}^{2}/(M_{Z^{\prime}Z^{\prime}}^{2}-M_{ZZ}^{2})<10^{-3},
\label{zmix}
\end{equation}
and the bound on the heavy $Z^{\prime}$ mass \cite{Zplimit}
\begin{equation}
M_{Z^{\prime}}>(500-800)~GeV,  \label{zpm}
\end{equation}
which implies serious constrains on the vacuum expectation value (vev) of
the singlet and the new $U(1)^{\prime}$ gauge coupling $g^{\prime}$.

However, both models have similar form for the scalar potential, where the
singlet can play the same role during the EWPT dynamics. In this type of
model\footnote{%
Similar remark holds for models with a singlet like \cite{sms,amin}.}, the
singlet vev within the wrong vacuum could be nonzero, i.e., $\left\langle
S\right\rangle =x\neq 0$, and therefore, is temperature dependant during the
EWPT dynamics. This feature could delay the EWPT, i.e., lowers the critical
temperature, and enhances the parameters that define the strong first-order
phase transition criterion \cite{SFOPT}:
\begin{equation}
\upsilon \left( T_{c}\right) /T_{c}>1,  \label{v/t}
\end{equation}%
where $T_{c}$ is the critical temperature and $\upsilon \left( T\right) $ is
the temperature dependent scalar vev.

In this work, we will investigate the possibility of getting a strong
first-order phase transition within the minimal gauge extension of the MSSM,
UMSSM without adding any new field beside the usual singlet.

This paper is organized as follows: in the second section, we give a brief
review of the UMSSM model, define the effective potential and discuss
different constraints on the parameters. After that, we discuss the EWPT
dynamics and show how to get a first-order phase transition. In the fourth
section, we discuss our numerical results. Finally, we summarize our
results. The different field-dependant masses used in the estimation of the
effective potential are given in Appendix A.

\section{The UMSSM model}

The $U(1)^{\prime }$-MSSM (or UMSSM) is based on the gauge group $%
G=SU(3)_{c}\times SU(2)_{L}\times U(1)_{Y}\times U(1)^{\prime }$ with the
couplings $g_{3}$, $g_{2}$, $g_{1}$ and $g^{\prime }$, respectively \cite%
{UMSSM} and the superpotential is given by
\begin{equation}
W=\lambda S\epsilon _{ij}H_{1}^{i}H_{2}^{j}+Y_{U}\epsilon
_{ij}Q^{i}U^{c}H_{2}^{j}+Y_{D}\epsilon
_{ij}Q^{i}D^{c}H_{1}^{j}+Y_{L}\epsilon _{ij}L^{i}E^{c}H_{1}^{j},
\end{equation}%
where $\epsilon _{ij}$ is the antisymmetry $2\times 2$ tensor, $Y_{U}$, $%
Y_{D}$ and $Y_{L}$ are Yukawa couplings, and $\lambda $\ is a coupling
constant in which $\lambda \left\langle S\right\rangle $ generates the $\mu $%
-term in the MSSM. The particle content of this model is given by the
left-handed chiral superfields $L\sim (1,2,-1/2,Q_{L})$, $E^{c}\sim
(1,1,1,Q_{E})$, $Q_{i}\sim (3,2,1/6,Q_{Q})$, $U^{c}\sim (\bar{3}%
,1,-2/3,Q_{U})$, $D^{c}\sim (\bar{3},1,1/3,Q_{D})$, $H_{1}\sim
(1,2,-1/2,Q_{1})$, $H_{2}\sim (1,2,1/2,Q_{2})$ and $S\sim (1,1,0,Q_{S})$,
where the $U(1)^{\prime }$ charges, $Q$'s, are model dependent. For
instance, in a class of $E_{6}$ gauge models, the group can be broken in two
steps to its $SO(10)$ and $SU(5)$ subgroups:

\begin{equation*}
E_{6}\rightarrow SO(10)\times U(1)_{\psi}\rightarrow SU(5)\times U(1)_{\psi
}\times U(1)_{\chi},
\end{equation*}
and in this case, the $U(1)^{\prime}$ generator is given in terms of the $%
U(1)_{\psi}$ and $U(1)_{\chi}$ generators and mixing angle $\theta_{E_{6}}$
\cite{UMSSM}

\begin{equation}
Q^{\prime }=Q_{\psi }\cos \theta _{E_{6}}+Q_{\chi }\sin \theta _{E_{6}}.
\end{equation}%
Although the $SU(3)_{C}\times SU(2)_{L}\times U(1)_{Y}\times U(1)^{\prime }$
model is motivated in the $E_{6}$ framework, we will not single out a
particular charge pattern and we only require that the model be anomaly
free. However, with the above particle content, one can easily show that the
invariance of the Yukawa terms in the superpotential under $U(1)^{\prime }$
and the absence of the $SU(3)_{C}-SU(3)_{C}-U(1)^{\prime }$ anomaly implies
that $Q_{s}=0$, and the $\mu $ problem arises again. Thus, anomaly
cancelation requires exotic representations beyond those of the MSSM. The
simplest extension for the anomaly to vanish is to assume three generations
of heavy (few TeV) vectorlike pairs of chiral fields $K_{i}$ and $K_{i}^{c}$
which transform as $(3,1)$ and $(\overline{3},1)$ under $SU(3)_{C}\times
SU(2)_{L}$ and opposite hypercharges \cite{AFREE}.

\subsection{The effective potential}

In case where both squarks or/and sneutrinos do not develop vevs, the scalar
potential is a combination of the so-called $D$, $F$ and soft terms, which
are given by
\begin{align}
V_{D}& =\frac{g_{2}^{2}+g_{1}^{2}}{8}\left(
H_{2}^{+}H_{2}-H_{1}^{+}H_{1}\right) ^{2}+\frac{g_{2}^{2}}{2}\left\vert
H_{1}^{+}H_{2}\right\vert ^{2}+\frac{g^{\prime 2}}{2}\left\vert
Q_{1}H_{1}^{+}H_{1}+Q_{2}H_{2}^{+}H_{2}+Q_{S}\left\vert S\right\vert
^{2}\right\vert ^{2},  \notag \\
V_{F}& =\left\vert \lambda \right\vert ^{2}\{\left\vert \epsilon
_{ij}H_{1}^{i}H_{2}^{j}\right\vert ^{2}+\left\vert S\right\vert ^{2}\left[
H_{1}^{+}H_{1}+H_{2}^{+}H_{2}\right] \},  \notag \\
V_{soft}&
=m_{1}^{2}H_{1}^{+}H_{1}+m_{2}^{2}H_{2}^{+}H_{2}+m_{S}^{2}\left\vert
S\right\vert ^{2}+\{A_{\lambda }S\epsilon _{ij}H_{1}^{i}H_{2}^{j}+h.c\}.
\label{tree}
\end{align}%
Here $m_{1}^{2}$,~$m_{2}^{2}$,~$m_{S}^{2}$ and $A_{\lambda }$\ are
usually called the SUSY soft parameters. The charges $Q^{\prime }$s
should be chosen in such a way that the anomaly cancelations\ are
ensured \cite{UMSSM,UDM}.

The structure of the tree-level potential (\ref{tree}) seems to allow two
explicit relative $CP$ violating phases between the doublets and the
singlet. However, gauge invariance dictates that the only allowed phase in
the potential, is that of the combination $S\epsilon _{ij}H_{1}^{i}H_{2}^{j}$%
, since gauge rotations can be used to set the charged Higgs vev $%
\left\langle H_{1}^{+}\right\rangle =0$, and the condition $\left\langle
H_{2}^{-}\right\rangle =0$\ implies that the physical charged Higgs is
nontachyonic ($M_{H^{\pm }}^{2}>0$) \cite{UMSSM}. Therefore, the general
form of the ground state could be written as%
\begin{equation}
\left\langle H_{1}\right\rangle =\frac{\upsilon _{1}}{\sqrt{2}}\left(
\begin{array}{c}
0 \\
1%
\end{array}%
\right) ,~\left\langle H_{2}\right\rangle =\frac{\upsilon _{2}}{\sqrt{2}}%
\left(
\begin{array}{c}
1 \\
0%
\end{array}%
\right) ,~\left\langle S\right\rangle =\frac{\upsilon _{x}e^{i\theta }}{%
\sqrt{2}}.  \label{back}
\end{equation}%
Here $\upsilon =(\upsilon _{1}^{2}+\upsilon _{2}^{2})^{1/2}=246~GeV$, and $%
\theta $ is the relative $CP$ violating phase between the singlet and the
doublets; and $\theta _{0}$ its value at the ground state. The tree-level
scalar potential in terms of the neutral components is given by%
\begin{align}
V\left( h_{1},h_{2},S\right) & =\left[ \left( g_{1}^{2}+g_{2}^{2}\right)
/4+g^{\prime 2}Q_{1}^{2}\right] h_{1}^{4}/8+\left[ \left(
g_{1}^{2}+g_{2}^{2}\right) /4+g^{\prime 2}Q_{2}^{2}\right]
h_{2}^{4}/8+g^{\prime 2}Q_{S}^{2}S^{4}/8  \notag \\
& +\left[ \lambda ^{2}-\left( g_{1}^{2}+g_{2}^{2}\right) /4+g^{\prime
2}Q_{1}Q_{2}\right] h_{1}^{2}h_{2}^{2}/4+\left[ \lambda ^{2}+g^{\prime
2}Q_{1}Q_{S}\right] h_{1}^{2}S^{2}/4  \notag \\
& +\left[ \lambda ^{2}+g^{\prime 2}Q_{2}Q_{S}\right]
h_{2}^{2}S^{2}/4+m_{1}^{2}h_{1}^{2}/2+m_{2}^{2}h_{2}^{2}/2+m_{S}^{2}S^{2}/2+%
\{A_{\lambda }Sh_{1}h_{2}e^{i\delta }/2\sqrt{2}+h.c\}.  \label{Vv}
\end{align}%
The neutral scalar sector in this model contains three $CP$-even scalars and
one $CP$-odd scalar. The tadpole minimization conditions at the ground state
along the $CP$-odd scalar forces the relative $CP$ phase $\theta $, to be
canceled by the phase of the parameter $A_{\lambda }$ in (\ref{tree}), which
must be taken as $A_{\lambda }=\pm \left\vert A_{\lambda }\right\vert
e^{-i\theta _{0}}$.\ This makes the ground state at tree-level independent
of this relative phase, and the phase $\delta $\ in (\ref{Vv}) is just $%
\delta =\theta -\theta _{0}$. It is clear that the dependence of (\ref{Vv})
on the phases disappears at the ground state. However, when considering the
one-loop corrections in the tadpole minimization condition along the $CP$%
-odd scalar, the tree-level phase cancelation is no longer valid at
one-loop, and the argument of $A_{\lambda }$ differs slightly than
$-\theta _{0}$ (or $-\theta _{0}+\pi $).

The structure of the tree-level potential (\ref{tree}) implies the presence
of a mixing between the submatrices ($CP$-odd scalars, Goldstone bosons, and
$CP$-even scalars) at tree-level for any values of the scalar fields except
at the ground state, however this mixing is proportional to $\sin \delta $.
This will ensure the $CP$ conservation at tree-level in the scalar sector.
Indeed, in addition to the phase appearance in the scalar sector due to the
deviation of Arg($A_{\lambda }$) from $-\theta _{0}$, the $CP$ violation
effect could be seen in this model at one-loop \cite{CP1}, and also through
the dependence on $\theta $ and not $\delta $\ can be seen in the
superpartner masses (\ref{stop}), (\ref{cha}) and (\ref{xi}).

The parameters $m_{1}^{2}$, $m_{2}^{2}$, and~$m_{S}^{2}$\ can be eliminated
by taking ($\upsilon _{1},\upsilon _{2},\upsilon _{x}e^{i\theta _{0}}$) as a
minimum of the effective potential. The vacuum stability of (\ref{Vv})
requires the condition:%
\begin{equation}
\lambda ^{4}-\lambda ^{2}\left[ \left( g_{1}^{2}+g_{2}^{2}\right)
/2+2g^{\prime 2}\left( Q_{S}^{2}-Q_{1}Q_{2}\right) \right] +\frac{3}{4}%
g^{\prime 2}\left( g_{1}^{2}+g_{2}^{2}\right) Q_{S}^{2}>0,  \label{Vsta}
\end{equation}%
where the $U(1)^{\prime }$ gauge invariance condition $Q_{1}+Q_{2}+Q_{S}=0$\
is taken into account.

In models that include singlets, like the present one, one needs to be
careful about whether the minimum ($\upsilon _{1},\upsilon _{2},\upsilon
_{x}e^{i\theta _{0}}$) is the absolute one. This can be checked by comparing
the effective potential value at this minimum with its value at the wrong
vacuum, which could be, in singlet models, ($0,0,x$) rather than the origin (%
$0,0,0$). The wrong vacuum can be defined as the minimum of the effective
potential in the direction where all the $SU(2)$ doublets vanish. At
tree-level, it is given by%
\begin{equation}
\left\vert x\right\vert ^{2}=\upsilon _{x}^{2}+\left[ Q_{1}/Q_{S}+\lambda
^{2}/g^{\prime 2}Q_{S}^{2}\right] \upsilon _{1}^{2}+\left[
Q_{2}/Q_{S}+\lambda ^{2}/g^{\prime 2}Q_{S}^{2}\right] \upsilon _{2}^{2}+%
\frac{\sqrt{2}}{g^{\prime 2}Q_{S}^{2}}\frac{A_{\lambda }}{\upsilon _{x}}%
\upsilon _{1}\upsilon _{2}.  \label{X0}
\end{equation}%
The value of $x$ in (\ref{X0}) should be relaxed by including the one-loop
corrections given below in Eq. (\ref{V1}). In the case where the right hand
side of (\ref{X0}) is positive, the wrong vacuum ($0,0,x$)\ does exist,
otherwise it should be ($0,0,0$). In both cases, ($\upsilon _{1},\upsilon
_{2},\upsilon _{x}e^{i\theta _{0}}$) must be the absolute minimum for the
effective potential. This condition imposes additional constraints on the
parameters of the model.

The one-loop effective potential at zero temperature is given in the $%
\overline{DR}$ scheme by \cite{NaLa}%
\begin{equation}
V^{1-l}\left( h_{1},h_{2},S\right) =V\left( h_{1},h_{2},S\right) +\sum_{i}%
\frac{n_{i}m_{i}^{4}}{64\pi ^{2}}\left( \log \frac{m_{i}^{2}}{\Lambda ^{2}}-%
\frac{3}{2}\right) ,  \label{V1}
\end{equation}%
where $m_{i}\left( h_{1},h_{2},S\right) $ are the field-dependant masses,
which are given in Appendix A, $\Lambda $\ is the renormalization scale,
which is taken to be $\Lambda =\upsilon =246~GeV,$ and $n_{i}$ are the
fields multiplicities: $n_{W}=6$, $n_{Z}=n_{Z^{\prime }}=3$, $%
n_{h_{1}^{0}}=n_{h_{2}^{0}}=n_{S}=n_{A^{0}}=n_{G}=1$, $n_{t}=-12$, $n_{%
\tilde{t}_{L}}=n_{\tilde{t}_{R}}=6$, $n_{\tilde{\chi}}=-2$, $n_{\tilde{C}}=-4
$, where $A^{0}$, $G$, $\tilde{t}_{L,R},$ $\tilde{\chi}$ and $\tilde{C}$
denote the $CP$-odd Higgs, Goldstone boson, left- and right-handed squarks,
neutralinos and charginos, respectively.

The thermal corrections to the effective potential can be computed using the
known techniques \cite{Th}. The one-loop effective potential at finite
temperature is given by%
\begin{equation}
V_{eff}\left( h_{1},h_{2},S,T\right) =V^{1-l}\left( h_{1},h_{2},S\right)
+T^{4}\sum_{i}n_{i}J_{B,F}\left( m_{i}^{2}\left( h_{1},h_{2},S\right)
/T^{2}\right) ;  \label{Vt}
\end{equation}
with%
\begin{equation}
J_{B,F}\left( \alpha\right) =\frac{1}{2\pi^{2}}\int_{0}^{\infty}x^{2}\log(1%
\mp\exp(-\sqrt{x^{2}+\alpha})),  \label{JBF}
\end{equation}
$n_{i}$ are given above and $m_{i}^{2}\left( h_{1},h_{2},S\right) $ are
given in Appendix A. In this work, to take into account all the heavy and
light fields, we will evaluate this integral numerically.

\subsection{The parameters}

In this model, we have many parameters, some of which are free like: $%
g^{\prime }$, $\lambda $, $\upsilon _{x}$, $\tan \beta =\upsilon
_{1}/\upsilon _{2}$, and the soft terms: $m_{Q}$, $m_{U}$, $A_{t}$, $%
A_{\lambda }$, $M_{2}$, $M_{1}$, and $M_{1}^{\prime }$; and others that are
fixed by a measured physical quantities such as : $g_{1}$, $g_{2}$, $%
\upsilon $ and $y_{t}$; or can be conditions like the elimination of $%
m_{1}^{2}$, $m_{2}^{2}$, and $m_{S}^{2}$ in (\ref{Vv}).

In scanning the parameter space of this model, we take into account the
constraint $Q_{1}+Q_{2}+Q_{s}=0$, conditions from the minimization of the
potential, the perturbativity of the quartic couplings in (\ref{Vv}), and
the vacuum stability (\ref{Vsta}). Another constraint could be derived from
the upper bound on the mixing between the gauge boson $Z$ and the new one $%
Z^{\prime }$ (\ref{zmix}), and the lower bound on the $Z^{\prime }$ mass (%
\ref{zpm}). The condition (\ref{zpm}), could be achieved by considering
relatively large $\upsilon _{x}$ or large $\left( g^{\prime }Q^{\prime
}\right) $. The condition (\ref{zmix}) could be fulfilled if the mixing term
$M_{ZZ^{\prime }}^{2}$ is vanishing, i.e.,%
\begin{equation}
Q_{1}=Q_{2}\tan ^{2}\beta ,  \label{zmixing}
\end{equation}%
which leads to a fine tuning in the values of $Q_{1,2}$ and $tan\beta $. The
second possibility is making $M_{Z^{\prime }Z^{\prime }}^{2}>>M_{ZZ}^{2},\
M_{ZZ^{\prime }}^{2}\,$, which roughly means%
\begin{equation}
g^{\prime }\left\vert Q_{S}\right\vert {\upsilon _{x}}\gtrsim (500-800)~GeV.
\label{supmixing}
\end{equation}

In our search for the parameter's space that fulfills the strong first-order
phase transition criterion, $\upsilon \left( T_{c}\right) /T_{c}>1$, we will
focus on the two following regions :

\textbf{(1)} Moderate values for the parameters $Q_{1,2}$ and $tan\beta $,
where (\ref{zmix}) is nearly satisfied. In this case, the singlet vev $%
\upsilon _{x}$ can be of order $\upsilon $ or relatively smaller.

\textbf{(2)} The two terms $M_{ZZ}^{2}$ and $M_{ZZ^{\prime }}^{2}$ in (\ref%
{Mzz}), are suppressed with respect to the mass term $M_{Z^{\prime
}Z^{\prime }}^{2}$. In this case, the values of $U^{\prime }(1)$ charge and
the vev of the singlet, $Q_{s}$ and $\upsilon _{x}$, must be large enough.

We should also distinguish between the two cases where the minimum ($%
0,0,x\neq 0$) of the potential does exist or not. The second case could be
ensured by choosing the parameter $A_{\lambda }$ in (\ref{X0}) to be
extremely negative, then $x$ does not exist. This is easier to satisfy in
region \textbf{(1)}. In both cases, the condition (\ref{zpm}) is, almost,
automatically fulfilled within (\ref{zmix}).

Also, one needs to know the effect of the phase $\theta$ on the EWPT
strength. At tree-level (\ref{Vv}), this phase is not relevant due to the
choice of $A_{\lambda}$, but the thermal corrections depend on this phase
through all the fields masses expect gauge bosons and top quark. Therefore,
we consider the values $0<\theta_{0}<\pi/3$.

The mass parameters: $m_{Q}$, $m_{U}$, $A_{t}$, $M_{2}$, $M_{1}$, and $%
M_{1}^{\prime }$ appear at one-loop level in the effective potential,
therefore we expect that their role is less important in the EWPT dynamics.
Indeed, these parameters will change the field\ masses and therefore could
make ($\upsilon _{1},\upsilon _{2},\upsilon _{x}e^{i\theta _{0}}$) a local
minimum instead of the global one through the corrections in (\ref{V1}).
However, the parameters: $g^{\prime }$, $\lambda $, $\upsilon _{x}$ and $%
\tan \beta $, as well the charges $Q^{\prime }$s that appear multiplied by $%
g^{\prime }$, seem to play a very important role for the strength of the
EWPT. Therefore we focus on these parameters while fixing $g^{\prime
}=g_{1},~m_{Q}=m_{U}=1~TeV,$ and choosing different values for $A_{t},\
M_{2},\ M_{1}$, and $M_{1}^{\prime }$. We will allow the parameters $\lambda
,\ \upsilon _{x},\ Q_{1,2},\ A_{\lambda }$, and $\theta _{0}$ to vary
randomly within the intervals :%
\begin{equation}
\begin{array}{ccc}
0.001<\lambda <0.5, &  & 0.5<\upsilon _{x}/TeV<4, \\
1<\tan \beta <20, &  & -1.2<A_{\lambda }/TeV<1.2, \\
-4<Q_{1,2}<4, &  & 0<\theta _{0}<\pi /3.%
\end{array}
\label{pra}
\end{equation}

\section{The electroweak phase transition}

Due the condition (\ref{zmix}) and (\ref{zpm}), there could exist a
hierarchy between the vev of the singlet and those of the doublets, i.e., $%
\upsilon _{x}>>\upsilon _{1,2}$. In this case, the gauge symmetry could be
broken in two step. However, in the case where the mixing is extremely
suppressed as in region \textbf{(1)} (\ref{zmixing}), the singlet vev could
be low as $\sim 500$ $GeV$. In this case, the gauge symmetry could be broken
just in one step. In the two-step symmetry breaking case, one notices that
above a certain high temperature, the singlet vev was zero $\upsilon
_{x}\left( T_{>>}\right) =0$. This can be seen by putting $\upsilon _{1,2}=0$
in (\ref{V1}), and taking only the thermal correction of $Z^{\prime }$. Then
at lower temperatures, it is not sure that the system moves directly from $%
(0,0,0)$ to ($\upsilon _{1},\upsilon _{2},\upsilon _{x}e^{i\theta _{0}}$),
or via an intermediate step: $(0,0,0)\overset{T_{c}^{\prime }}{\rightarrow }%
(0,0,x)\overset{T_{c}}{\rightarrow }(\upsilon _{1},\upsilon _{2},\upsilon
_{x}e^{i\theta _{0}})$. This will depend, in general, on the theory
parameters, especially the value of the singlet vev. If it is comparable to
the EW vev $\upsilon $, the phase transition could occur once, however if it
is much larger than $\upsilon $, the phase transition will occur in two
steps.

Since the singlet dynamics does not affect the $SU(2)$ sphaleron processes,
we will not be interested in distinguishing between the one- and two-steps
symmetry breaking. We will treat our field dynamics using the effective
potential where the singlet is replaced by a temperature-dependant vev,
i.e.,
\begin{equation}
\mathcal{V}\left( h_{1},h_{2},T\right) =V_{eff}\left(
h_{1},h_{2},x(T),T\right) .
\end{equation}%
At higher temperatures, the effective potential admits only one minimum
where $\upsilon _{1,2}\left( T\right) =0$. As the Universe cools down, the
effective potential acquires a new minimum $\upsilon _{1,2}\left( T\right)
\neq 0$; but\ it is not the absolute one. At such temperature, the critical
temperature $T_{c}$, the two minima get degenerate%
\begin{equation}
V_{eff}\left( \upsilon _{1}^{c},\upsilon _{2}^{c},\upsilon
_{x}^{c},T_{c}\right) =V_{eff}\left( 0,0,x^{c},T_{c}\right) .
\end{equation}%
Below this temperature, the new minimum becomes the absolute one, and the
system has to move from the old (false) vacuum to the new (true) one. In the
case where a barrier does exist between the two minima, this transition has
to occur via tunneling trough bubbles nucleation at certain points, which
expand and the whole space by the new vacuum $\upsilon _{1,2}\left( T\right)
\neq 0$, i.e., the symmetry is broken.

If the effective potential (\ref{Vt}) is expanded as powers of $m/T$ (in the
limit $m<<T)$ in a similar way to the SM in the doublets direction, the two
leading terms are of order $h^{2}T^{2}$ and $h^{3}T$. The first term
determines the temperature when the barrier between the potential two minima
disappears, while the second term is relevant to the strength of the
first-order phase transition. It turns out that a resummation of the
so-called Daisy diagrams \cite{daisy}, leads to a contribution of order $%
h^{3}T$ which exactly cancels the contribution from certain particles (e.g.
Higgs and longitudinal gauge bosons). Therefore this screening effect could
weaken the strength of the EWPT. For that, we will check the importance of
this effect by using a resumed effective potential where (\ref{Vt}) is
modified by replacing the bosonic field-dependant masses by their thermally
corrected values \cite{DWG}. The thermal corrections to the bosonic masses
are given in Appendix B.

The $B+L$ anomalous interactions \cite{B+L} that violate the baryon number
do not have the same rate in the symmetric and broken phases ( i.e., at both
sides of the bubble wall). In the symmetric phase, this rate behaves like $%
\sim T^{4}$ \cite{Symm}, and is suppressed as $exp(-E_{Sp}/T)$ \cite{bro},
in the broken phase, where $E_{Sp}$ is the system static energy within such
field configuration called the sphaleron \cite{Sph}. Therefore any generated
baryon number at the symmetric phase will get erased , unless these
interactions are switched off in the broken phase, which translates to the
condition (\ref{v/t}). In reality, the $B+L$ anomalous interactions should
be switched off at a temperature $T_{0}<T_{n}<T_{c}$, where $T_{n}$ is the
temperature at which bubbles start to nucleate, and $T_{0}$ is the
temperature at which the barrier between the two minima completely
disappears. Then the condition in (\ref{v/t}) should be fulfilled at $T_{n}$%
, but, in general, the two values are significantly close, and if $v(T_{c})/{%
T_{c}}>1$, it is necessarily satisfied at $T_{n}$. Indeed, in some earlier
works (e.g. \cite{T0,MSSMcheck}), the condition (\ref{v/t}) at $T_{0}$ was
used in defining the phase transition. This means that the anomalous $B+L$
violating processes should be switched off at the end of the phase
transition, i.e., the Universe is filled by the true vacuum by the expanded
bubbles. The value $T_{0}$, usually called the lower metastability
temperature or lower spinodial decomposition point \cite{MSSMcheck}, can be
determined using the Jacobian $det[\partial ^{2}V(h,\ T_{0})/\partial
h_{i}\partial h_{j}]_{h=0}=0$. However, in our work, it is safer to consider
the condition (\ref{v/t}) at $T_{c}$, that is defined in (\ref{TC}).

Since the singlet field does not play an important role in the sphaleron
processes \cite{amin}, the criterion of a strong first-order phase
transition in our case is given by%
\begin{equation}
\upsilon \left( T_{c}\right) /T_{c}\equiv \sqrt{\upsilon _{1}^{2}\left(
T_{c}\right) +\upsilon _{2}^{2}\left( T_{c}\right) }/T_{c}>1.  \label{cri}
\end{equation}

In the general case where the relative phase $\theta \neq 0$, the field
ground state at nonzero temperature should be written as $\left\{
v_{i}\right\} _{i=1,4}$= ($\upsilon _{1}$, $\upsilon _{2}$, $\upsilon
_{x}\cos \theta $, $\upsilon _{x}\sin \theta $)\ instead of ($\upsilon
_{1},\upsilon _{2},\upsilon _{x}$) and the relative phase. These 4 variables
should be treated independently when looking for $\upsilon _{1,2,x}$\ and $%
\theta $\ at any temperature $T$. Then the phase transition could be defined
through the equations%
\begin{equation}
\frac{\partial }{\partial v_{i}}V_{eff}\left( v_{i},T_{c}\right)
=0,~V_{eff}\left( v_{1},v_{2},v_{3},v_{4},T_{c}\right) =V_{eff}\left(
0,0,x_{1},x_{2},T_{c}\right) ,  \label{TC}
\end{equation}%
where $x_{1,2}$\ are the real and imaginary parts of the $T$-dependent $x$
given in (\ref{X0}).

\section{Numerical results}

\begin{figure}[h]
\begin{center}
\includegraphics[width=8.5cm,height=6.5cm]{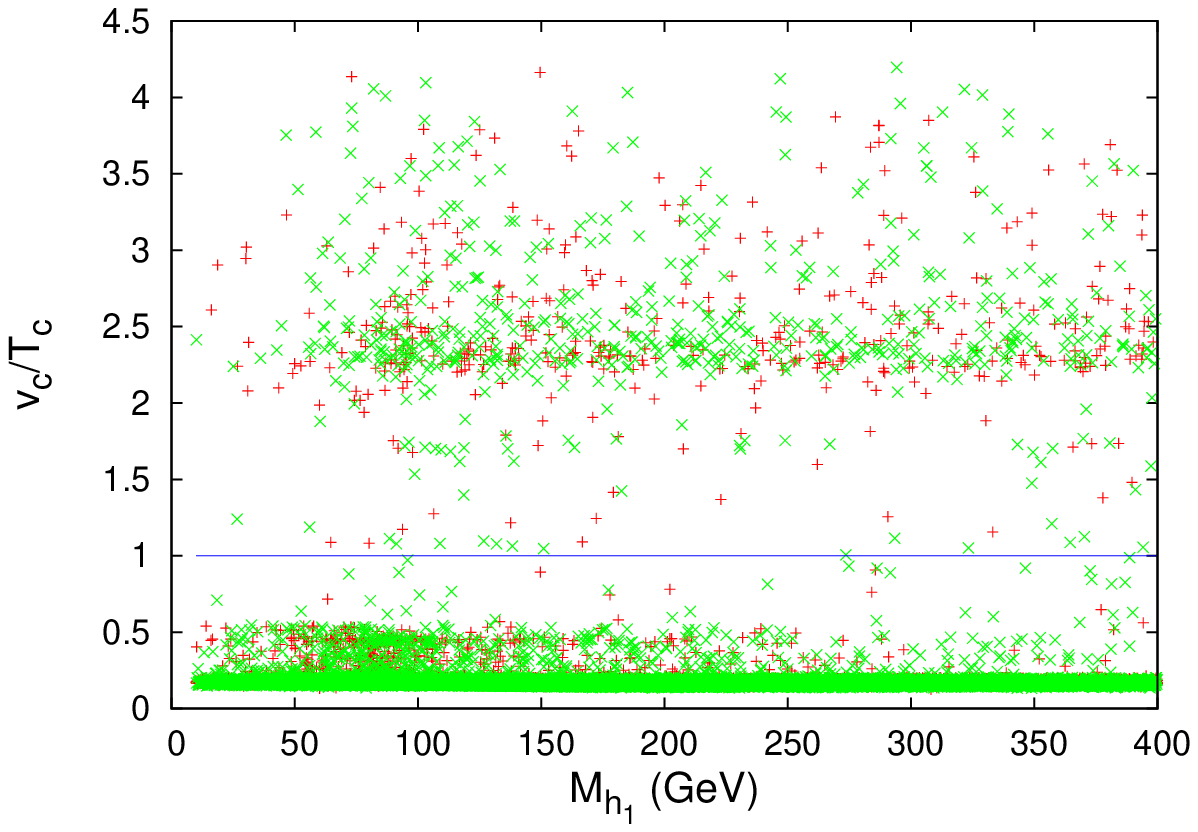}~~%
\includegraphics[width=8.5cm,height=6.5cm]{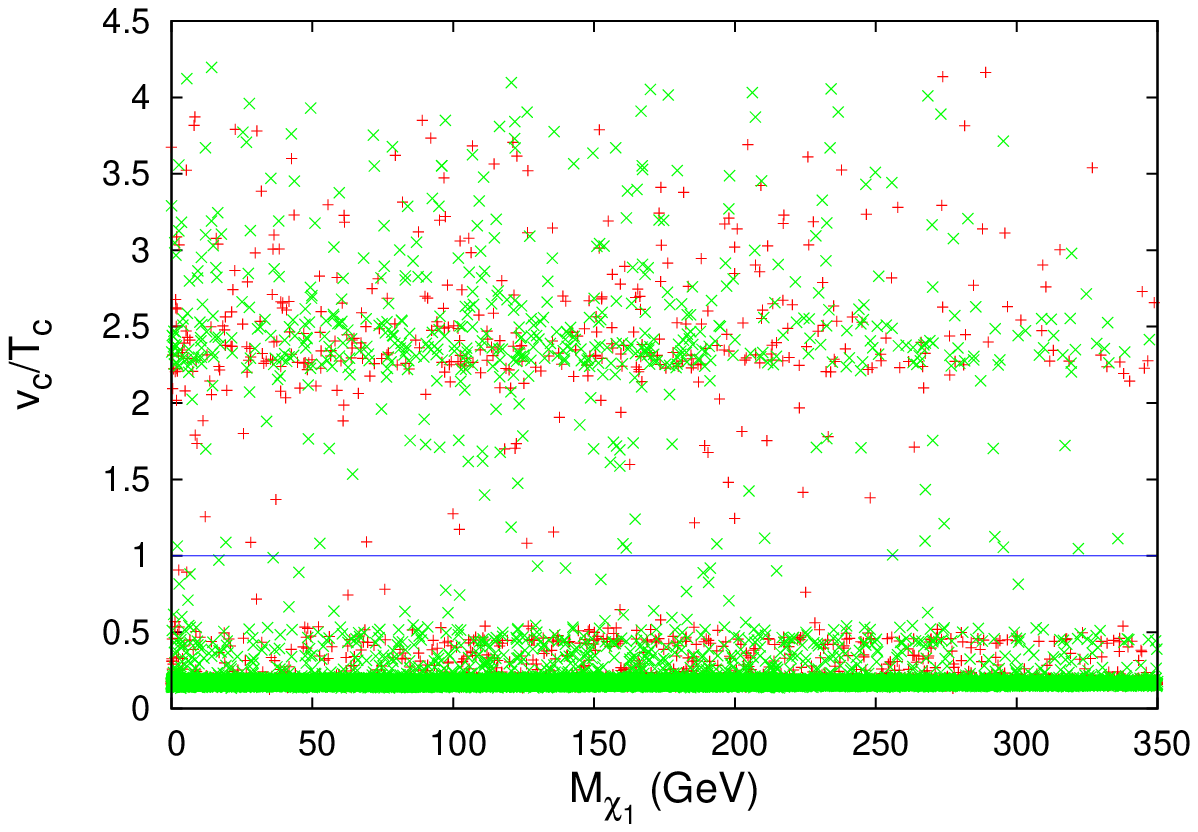}\\[0pt]
\includegraphics[width=8.5cm,height=6.5cm]{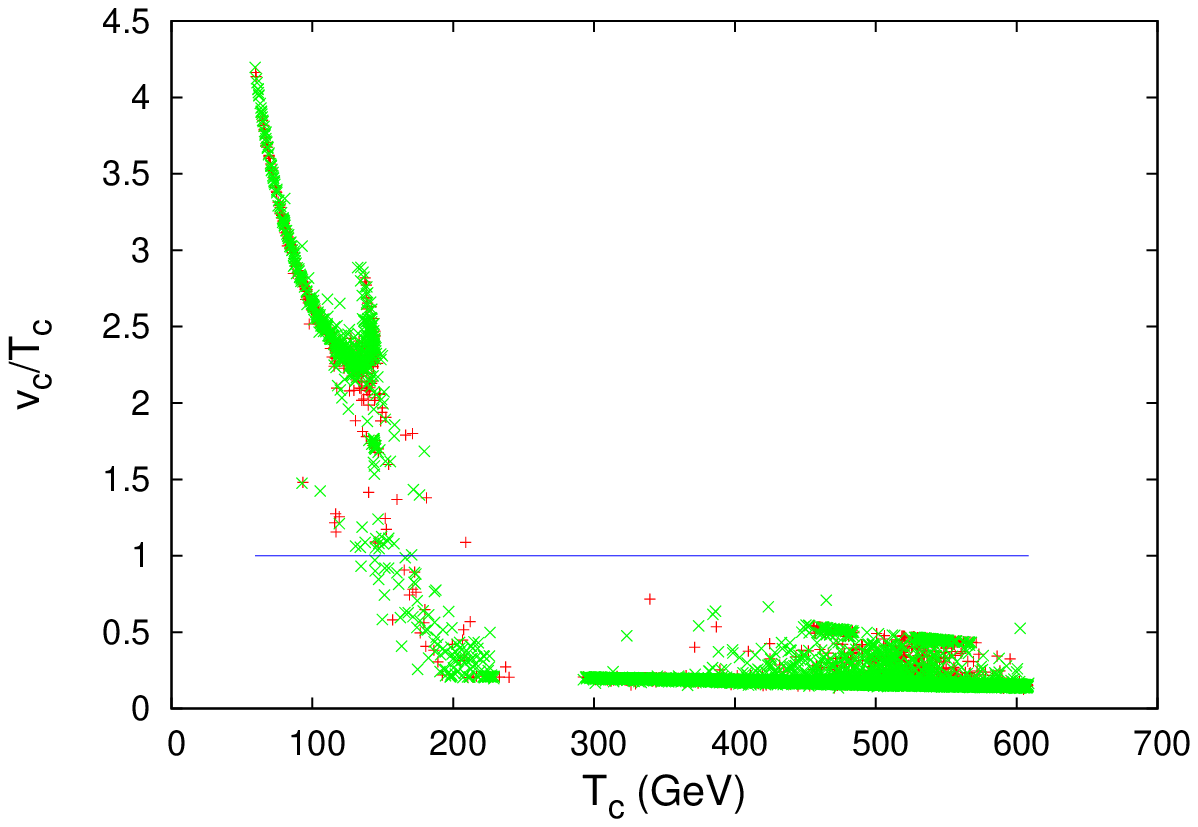}
\end{center}
\caption{\textit{The dependence of the quantity $\protect\upsilon _{c}/T_{c}$
on the lightest Higgs mass, on the lightest neutralino mass, and the
corresponding critical temperature, taking into account different values of
the parameters in (\protect\ref{pra}) is shown. The red points refer to
benchmarks from region \textbf{(1)}, and the green ones to benchmarks from
region \textbf{(2).}}}
\label{Mhn}
\end{figure}

In the following figures, we show the dependence of the quantity $\upsilon
\left( T_{c}\right) /T_{c}$ in (\ref{cri}) on the lightest Higgs mass (in
Fig. \ref{Mhn} left), and on the lightest neutralino mass (in Fig. \ref{Mhn}
right), for a random choice of about $10^{6}$ cases in both regions \textbf{%
(1)} and \textbf{(2)}. We find that only about 8\% of the benchmarks fulfill
the required conditions, and only 2.5\% of the survived benchmarks give a
strong first-order phase transition.

As it is clear from Fig. \ref{Mhn}, the EWPT could be strongly first-order
in the two regions \textbf{(1)} and \textbf{(2)}, for a large range of the
of the values of the lightest Higgs mass, and lightest neutralino mass.
These masses are estimated at tree-level, which are expected to be boosted
to larger values when considering one-loop corrections. One remarks that the
phase transition strength does depend on the lightest Higgs and neutralino
masses. One also remarks that the majority of the benchmarks that give a
weak phase transition have the value of $\upsilon \left( T_{c}\right)
/T_{c}\sim 0.14-0.18$, and their corresponding critical temperature is
between $T_{c}\sim 300-600$ GeV. While all the corresponding values of $%
T_{c} $ to the strong phase transition are $T_{c}\leq 180$ GeV.

The mixing with the singlet will modify the doublets couplings to the
fermions, and since the lightest Higgs contains a singlet amount, its mass
could be (much) smaller than SM bound $\sim $114 GeV. One remarks from Fig. %
\ref{Mhn} (right), that the EWPT could be strongly first-order for small
values of the lightest neutralino masses; $m_{\chi _{1}}<15$ GeV. Such a
light mass of neutralino with a spin-independent cross section of order $%
10^{-5}$pb could be a possible interpretation of the recent observations of
CoGeNT and DAMA/LIBRA \cite{LDM}. In particular, the lightest neutralino can
have a non negligible component of the superpartner of the $U(1)^{\prime }$
gauge boson \cite{next}.

The critical temperature is, in general, higher when comparing with minimal
SM ($\sim $100 $GeV$), the generic value is larger than $300-600$ $GeV$.
This is a consequence of the interaction of the doublets with the singlet
that has, in general, a very large vev. However for the benchmarks, giving a
strong first-order EWPT, $T_{c}$ is relatively smaller than the generic
value. In fact, for points \textbf{(1)} as $T_{c}$ is smaller than about $180
$ GeV, the stronger the EWPT becomes.

In order to understand this point, we take a benchmark from Fig. \ref{Mhn},
and study the dependence of the scalar vevs on the temperature $T$ below and
just above the critical temperature. We will also check \ how could this
behavior be changed with respect to the charges $Q^{\prime }$s, and other
parameters like $A_{t}$, and $M_{1}^{\prime }$, that appear in the effective
potential at one-loop level. Therefore we consider the benchmarks in Table-%
\ref{Tab}.
\begin{table}[h]
\begin{center}
\begin{tabular}{c|c|c|c|c|}
\cline{2-5}
& (a) & (b) & (c) & (d) \\ \hline
\multicolumn{1}{|c|}{$\lambda$} & 0.0235 & 0.0235 & 0.0235 & 0.0235 \\ \hline
\multicolumn{1}{|c|}{$tan\beta$} & 2.0566 & 2.0566 & 2.0566 & 2.0566 \\
\hline
\multicolumn{1}{|c|}{$A_{\lambda}$} & -208.9569 & -208.9569 & -208.9569 &
-208.9569 \\ \hline
\multicolumn{1}{|c|}{$\upsilon_{x}$} & 1173.3560 & 1173.3560 & 1173.3560 &
1173.3560 \\ \hline
\multicolumn{1}{|c|}{$Q_{1}$} & 0.3575 & 1.75 & 0.3575 & 0.3575 \\ \hline
\multicolumn{1}{|c|}{$Q_{2}$} & 1.6147 & 1.5 & 1.6147 & 1.6147 \\ \hline
\multicolumn{1}{|c|}{$\theta_{0}$} & 0.5888 & 0.5888 & 0.5888 & 0.5888 \\
\hline
\multicolumn{1}{|c|}{$M_{1}$} & 857.9520 & 857.9520 & 857.9520 & 1500 \\
\hline
\multicolumn{1}{|c|}{$M_{2}$} & 398.7435 & 398.7435 & 398.7435 & 398.7435 \\
\hline
\multicolumn{1}{|c|}{$M_{1}^{\prime}$} & 881.3455 & 881.3455 & 881.3455 &
881.3455 \\ \hline
\multicolumn{1}{|c|}{$A_{t}$} & 274.4727 & 274.4727 & 1400 & 274.4727 \\
\hline
\multicolumn{1}{|c|}{$T_{c}$} & 103.6387 & 326.2436 & 302.9225 & 100.8061 \\
\hline
\multicolumn{1}{|c|}{$\upsilon_{c}/T_{c}$} & 2.5836 & 0.1940 & 0.1976 &
2.6356 \\ \hline
\multicolumn{1}{|c|}{$m_{h_{1}}$} & 172.1886 & 163.3732 & 173.0714 & 172.1589
\\ \hline
\multicolumn{1}{|c|}{$m_{\chi_{1}}$} & 5.6907 & 5.6569 & 5.6907 & 6.2573 \\
\hline
\end{tabular}%
\end{center}
\caption{\textit{The values of the parameters used to study the scalar vevs
dependence with respect to the temperature are shown. The mass-dimension
parameters are given in GeV.}}
\label{Tab}
\end{table}

From this table, it is clear that the EWPT strength is extremely sensitive
to the charges $Q^{\prime }$s which represent in a way the strength of the $%
U(1)^{\prime }$ interactions since the $Q^{\prime }$ charges always appear
multiplied by $g^{\prime }$. However, the soft parameter $A_{t}$ and M1 that
appear in the effective potential at one loop could have an important effect
on the strength of the EWPT for some particular choices of the parameter
space. The dependence of the EWPT strength on the relative phase $\theta _{0}
$, could be seen by taking changing its value of the benchmark (a) in Table-%
\ref{Tab}. We find that its effect is extremely negligible [difference in
the value of $\upsilon \left( T_{c}\right) /T_{c}$ is $<$ 0.1\%].

\begin{figure}[h]
\begin{center}
\includegraphics[width=16cm,height=13cm]{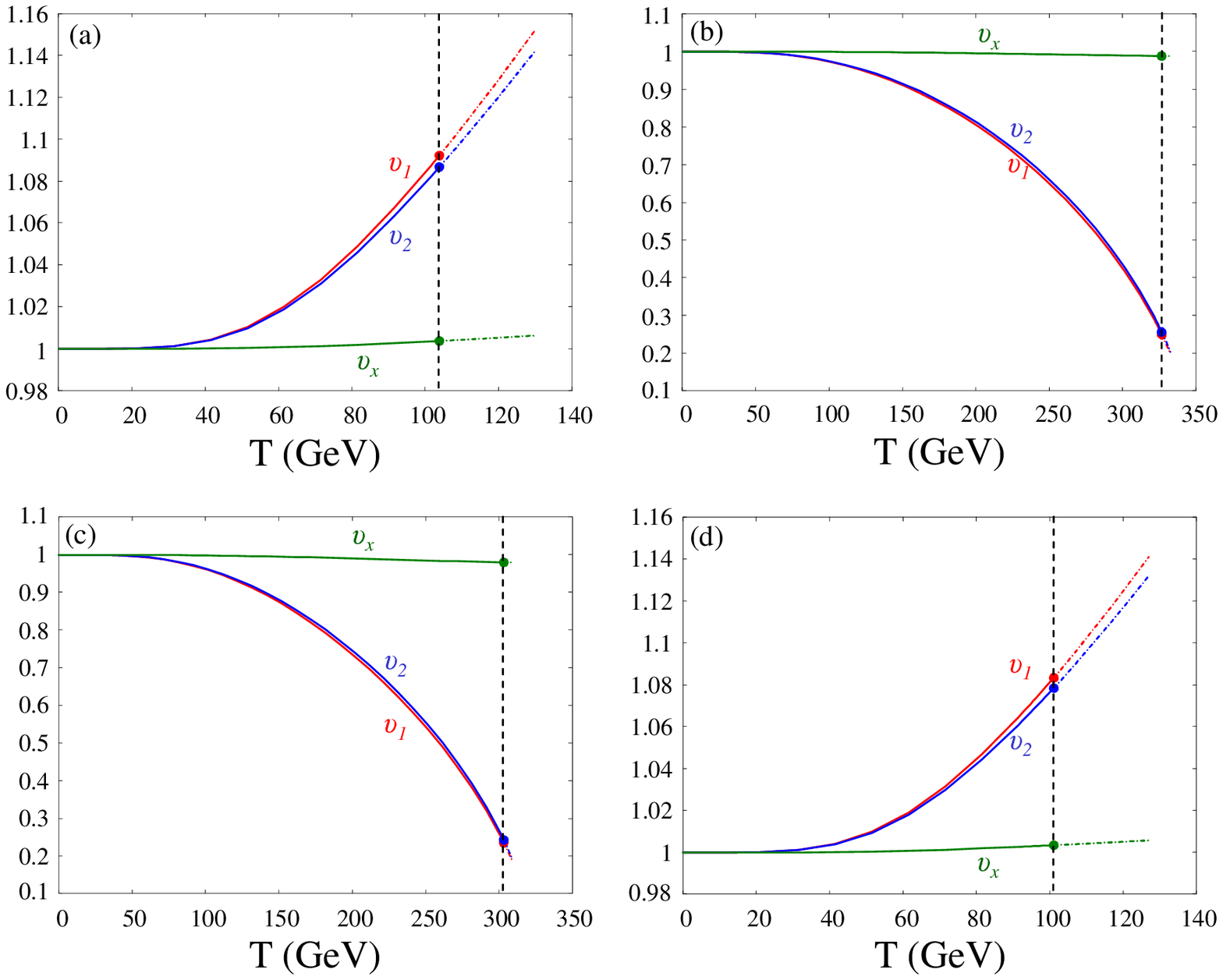}
\end{center}
\caption{\textit{The dependence of the scalar vevs on the temperature T is
shown. These quantities are scaled by their zero temperature values. The
solid curves refer to the broken phase, where (}$\protect\upsilon _{1},%
\protect\upsilon _{2},\protect\upsilon _{x}$\textit{) represents the global
minimum. While the dashed ones refer to the symmetric phase, where the
global minimum is (0,0,x(T)).}}
\label{cs}
\end{figure}
In Fig. \ref{cs}, we show the dependence of the ground state on the
temperature below the critical temperature for the benchmark (a) in Table-%
\ref{Tab}, and its modifications (b), (c) and (d). One remarks that the
common feature between all these cases is that the dependence of the singlet
vev on the temperature is very weak around and below the critical
temperature. This could be understood due to the fact that the singlet vev
is much larger than this temperature [$x(T<<x)\sim x(0)$], and also, due to
a possible two-stages phase transition realization, where the first stage $%
(0,0,0)\rightarrow (0,0,x(T))$, takes place around the temperature $T_{\ast
}\sim x$; therefore at lower temperatures $T<<T_{\ast }$, the singlet vev
becomes almost temperature independent.

Another important remark, is that the Higgs vevs for this benchmark [Fig. %
\ref{cs}-a] are increasing when the Universe gets cooled unlike the SM \cite%
{andhall}, or MSSM \cite{MSSMcheck}, and similar remark for (d). In cases
(b) and (c), the Higgs vevs decrease, but with slower pace than in the SM or
MSSM, which makes the critical temperature larger than in the SM. The
modified cases (b) and (c) belong to the majority of the benchmarks
mentioned before.

The effect of the increasing doublets' vev values at high temperatures with
respect to their zero temperature values has been mentioned in a similar
work \cite{ham}. This behavior and the slow decrease of the doublets vevs
with respect to the temperature, is a consequence of the interaction of the
singlet with the doublet Higgs. These interactions have the effect of
relaxing the shape of the potential in the direction of the doublets, and
therefore enhancing the ratio in (\ref{cri}). This is a common feature for
models with singlets \cite{amin}.

\begin{figure}[h]
\begin{center}
\includegraphics[width=16cm,height=6.5cm]{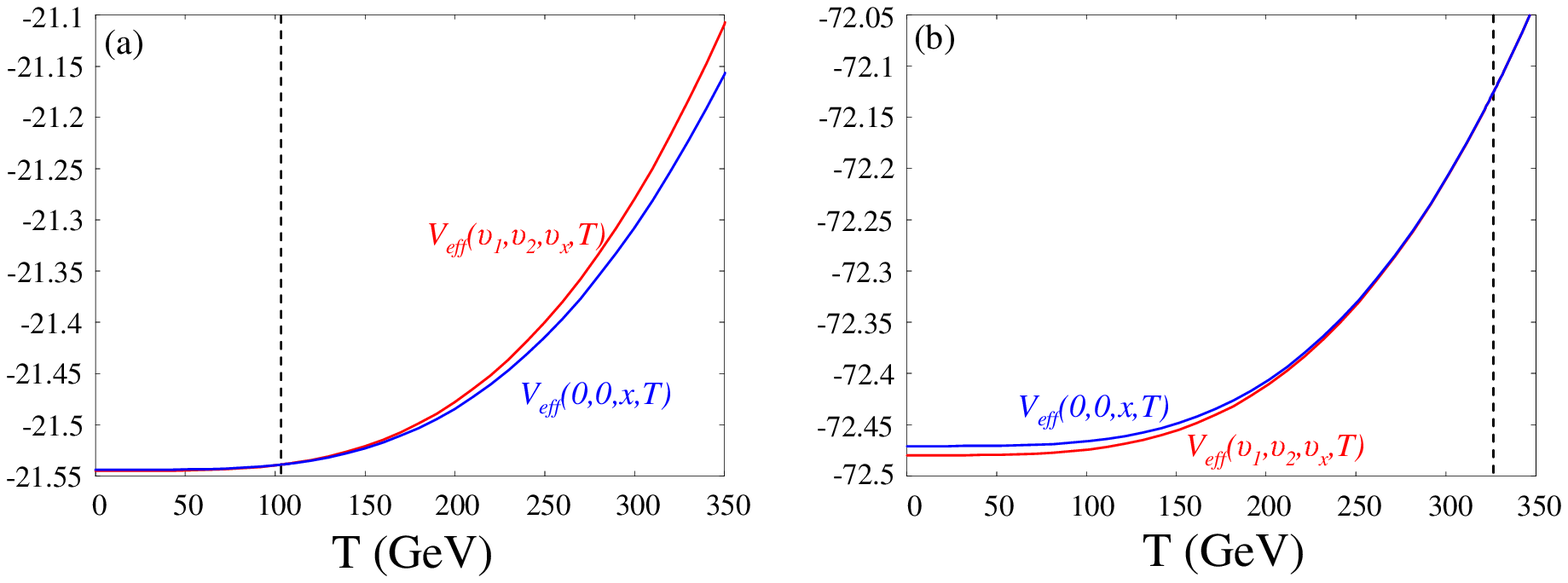}
\end{center}
\caption{\textit{The evolution of the effective potential values at the
minima (0,0,x) and (}$\protect\upsilon _{1}$\textit{,}$\protect\upsilon _{2}$%
\textit{,}$\protect\upsilon _{x}$\textit{) with respect to the temperature
T\ in units of }$\protect\upsilon ^{4}$\textit{\ is shown. In this
temperature range, the value of the effective potential at }$(0,0,0)$\textit{%
\ is about 5.879}$\sim $\textit{6.620 for benchmark (a), and 7.322}$\sim $%
\textit{8.131 for benchmark (b) in units of }$\protect\upsilon ^{4}$\textit{.%
}}
\label{vwr}
\end{figure}

From (\ref{TC}), the phase transition is defined as the degeneracy of the
two vacua. Unlike the SM and MSSM, the wrong vacuum $(0,0,x(T))$, is
evolving with respect to the temperature, and could be reached through a
phase transition at very high temperature. Therefore, its evolution could be
a very important factor which can strengthen the EWPT. It is defined from
the effective potential as the local minimum%
\begin{equation}
V^{r}\left( T\right) =V_{eff}(0,0,x(T),T)~~~,~~~\frac{\partial }{\partial S}%
V_{eff}(0,0,x(T),T)=0,  \label{Vr}
\end{equation}%
which was a global one before the phase transition, i.e., above the critical
temperature. We need to check that the evolution of (\ref{Vr}) is
responsible of lowering the critical temperature in (a) and (d), which does
not play the same role in (b) and (c). In Fig. \ref{vwr}, we show how does
the evolution of the wrong vacuum with respect to the temperature decrease
the critical temperature for the benchmarks (a) versus its modification (b).

From Fig. \ref{vwr}, one remarks that the minimum ($\upsilon _{1},\upsilon
_{2},\upsilon _{x}$) becomes the true one at low temperature for case (a),
and at large temperature for case (b) due to the evolution of the wrong
minimum ($0,0,x$). This can be identified from the intersection of the
thermal effective potential at ($\upsilon _{1}$,$\upsilon _{2}$,$\upsilon
_{x}$) and its value at ($0,0,x$). If one completely ignores the minimum ($%
0,0,x$), the EWPT, which is defined in this case by the intersection of this
value with effective potential at ($\upsilon _{1}$,$\upsilon _{2}$,$\upsilon
_{x}$) and its value at (0,0,0), will take place at very high temperatures,
normally of the order $\sim O(\upsilon _{x})$. It is clear that the wrong
vacuum evolution has the main role in lowering the critical temperature in
case (a) compared with case (b). It might seem that the EWPT is less
sensitive to some parameters such as $m_{Q},\ m_{U},$ $A_{t},\ M_{2},~M_{1}$%
, and $M_{1}^{\prime }$, since they appear in the effective potential at
one-loop. However, we have shown in Table-\ref{Tab} and in Fig. \ref{vwr}
that the EWPT dynamics is sensitive to these inputs more than expected. In
fact, if one adds a small perturbation to these parameters, then the wrong
vacuum, $(0,0,x)$, becomes the absolute one at zero temperature, through the
one-loop corrections in (\ref{V1}), which will rule out our benchmark.

In order to estimate the screening effect on the EWPT strength, we consider
the modified effective potential \cite{DWG} by replacing the longitudinal
gauge bosons and squarks masses in (\ref{Vt}) by their thermally corrected
expression (\ref{Thga}) and (\ref{Thsq}) in Appendix B. We ignore the scalar
contributions because they are less relevant to the EWPT dynamics. One can
distinguish two different types of the benchmarks in Fig. \ref{Mhn}; the
first variant: the benchmarks with increasing doublet vevs with respect to
the temperature [like the benchmark (a) in Fig. \ref{cs}], and the second
variant are the benchmarks with decaying doublet vevs with respect to the
temperature like most of the benchmarks in~Fig. \ref{Mhn}.

For the second variant, it is expected to have a similar behavior as in the
(MS)SM-like models where the screening effect weakens significantly the
strength of the EWPT. The benchmarks whose generic values of $\upsilon
_{c}/T_{c}$ in the interval $0.14-0.18$ get reduced to $0.10-0.12$, that is
about $30\%$ effect. In fact, even the benchmarks that give a strong
first-order EWPT but with (slowly) decaying doublet vevs with respect to the
temperature become weak EWPT benchmarks after including the daisy-diagrams
contribution.

Without including the daisy-diagrams, the first variant correspond to
benchmarks giving a strong first-order phase transition whose critical
temperatures are less than $T_{c}<180\sim 200$ GeV. Once we include the
daisy diagrams contribution, we find that the doublets vevs become sharply
increasing with respect to the temperature, and this results in a very low
critical temperature ($T_{c}\sim 32-36~GeV$), and very strong ($\upsilon
_{c}/T_{c}\sim 7-9$) EWPT. This unusual effect is due to the breakdown of
the approximation in which one neglects the $mT$ term in the expression of
the thermal mass. In other words, one should expect that at low temperatures
the thermal mass is approximately given by couplings $\times $ zero
temperature mass instead of the $T^{2}$ term \cite{daisy}. For perturbative
couplings, this is smaller than the zero temperature mass of the particle
and can be neglected. Thus, unlike the second variant, we expect that the
screening effects for the first variant to be negligible and do not
significantly reduce the strength of the phase transition found earlier
(which is first-order). To estimate how small the screening effect is,
requires taking into account the exact thermal corrections in the effective
potential which is beyond the scope of this paper.

It is worth noticing that large values of $\upsilon \left( T_{c}\right)
/T_{c}$ [i.e., larger than $\upsilon \left( T_{c}\right) /T_{c}>3.5$] can be
easily obtained. This corresponds to a severe suppression of the sphaleron ($%
B+L$ violating) processes inside the bubbles, and not necessary to the
freezing of the Universe in the wrong vacuum. The decay of the wrong vacuum
is related to the bubbles' dynamics, not to the (non-)efficiency of the $B+L$
violating processes. This point requires a special careful investigation to
put constraints on the theory parameters from the fact that the wrong vacuum
must decay into the true one.

\section{Conclusion}

In this work, the nature of the electroweak phase transition within the
minimal $U(1)$ extension of the MSSM (UMSSM) without including exotic
particles, has been investigated. We found that the EWPT could be strongly
first-order for a large range of the lightest Higgs and neutralino masses,
without adding extra singlet scalars or fermions. We evaluated the effective
potential at one-loop taking into account the whole particle spectrum, and
its temperature-dependant corrections were estimated exactly using the known
techniques.

We found that the strength of the EWPT could be enhanced due to two factors:
first, the interactions of the singlet scalars with the doublets which relax
the shape of the effective potential in the doublets directions, and
therefore, lead to a large value for the ratio $\upsilon \left( T\right) /T$
at the critical temperature. The second factor is that the
temperature-dependant local minimum, $(0,0,x\left( T\right) )$, could play
an important role during the EWPT dynamics. It can delay the phase
transition until relatively low temperatures (even below $100$ $GeV$), which
favor the ratio $\upsilon \left( T_{c}\right) /T_{c}$ to be large enough,
without conflicting the usual severe experimental constraints of the SM and
MSSM.

During this dynamics, the doublets vevs could be decaying with respect to
the temperature but slower than in the case of the SM or MSSM. Another
unusual behavior is that the doublets vevs could be increasing with respect
to the temperature, which leads to a very strong first-order EWPT. We found
that the inclusion of the ring contribution does weaken the EWPT if the
doublets vevs are decaying with respect to the temperature (the second
variant), and do not change significantly the strength of phase transition
in the opposite case.

We also mention that the strength of the EWPT, as well as the reliability of
the theory benchmarks, are very sensitive to the input parameters that
appear in the effective potential at one-loop.

\subsection*{Acknowledgements}

A. A. wants to thank the physics department at UAEU for hospitality during a
period where part of this work was performed, and S. N. wants to thank the
laboratory of theoretical physics at Jijel University for hospitality during
the last stage of this work. A. A is supported by the Algerian Ministry of
Higher Education and Scientific Research under the CNEPRU Project No.
D01720090023.

\appendix

\section{Field-dependent masses}

Here, we will present field-dependant masses with the existence of $CP$
phases; and by vanishing these phases, we get the $CP$ concerning case.

\textbf{Gauge bosons.} In this model, the gauge bosons masses do not depend
on $CP$ phases. They have similar masses as in the MSSM, but the $Z$ boson
is mixed with the new $Z^{\prime }$
\begin{align}
M_{W}^{2}& =\frac{1}{4}g_{2}^{2}\left( h{_{1}^{2}+}h{_{2}^{2}}\right) ,
\notag \\
M_{Z-Z^{\prime }}^{2}& =\left(
\begin{array}{cc}
\frac{1}{4}\left( g_{1}^{2}+g_{2}^{2}\right) \left( h{_{1}^{2}+}h{_{2}^{2}}%
\right)  & \frac{1}{2}g^{\prime }\sqrt{g_{1}^{2}+g_{2}^{2}}\left( Q_{1}h{%
_{1}^{2}}-Q_{2}h{_{2}^{2}}\right)  \\
\frac{1}{2}g^{\prime }\sqrt{g_{1}^{2}+g_{2}^{2}}\left( Q_{1}h{_{1}^{2}}%
-Q_{2}h{_{2}^{2}}\right)  & g^{\prime 2}\left( Q_{1}^{2}h{_{1}^{2}}%
+Q_{2}^{2}h{_{2}^{2}+}Q_{S}^{2}S{^{2}}\right)
\end{array}%
\right) .  \label{Mzz}
\end{align}

\textbf{Tops and Stops. }The stop masses are the same as in the MSSM with
replacing $\mu $\ but its effective value $\mu =\lambda S/\sqrt{2}$:

\begin{align}
m_{t}^{2} & =\frac{1}{2}y_{t}^{2}h_{2}^{2}  \notag \\
M_{\tilde{t}}^{2} & =\left(
\begin{array}{cc}
m_{Q}^{2}+m_{t}^{2}+(g_{2}^{2}-\frac{1}{3}{g_{1}^{2}})(h_{1}^{2}-h_{2}^{2})/8
& A_{t}h_{2}/\sqrt{2}+\lambda h_{1}Se^{i\theta}/2 \\
A_{t}h_{2}/\sqrt{2}+\lambda h_{1}Se^{-i\theta}/2 & m_{U}^{2}+m_{t}^{2}+{%
g_{1}^{2}}(h_{1}^{2}-h_{2}^{2})/6%
\end{array}
\right) .  \label{stop}
\end{align}

\textbf{Scalars. }The spectrum of physical Higgses after symmetry breaking
consists of three neutral $CP$ even scalars, a mixture between one $CP$ odd
pseudoscalar ($A^{0}$), and two Goldstone bosons that are absorbed by $Z$
and $Z^{\prime }$ respectively, and a mixture between a charged Higgs and
the Goldstone bosons that are absorbed by $W^{\pm }$. In the ground state $%
h_{i}=\left\langle h_{i}\right\rangle $, there is no mixture between the
real and imaginary parts of the complex fields of the scalar sector, and the
squared-mass matrix should be represented in two independent $3\times 3$
matrices in the two basis [$\sqrt{2}Re(H_{1}^{0})$, $\sqrt{2}Re(H_{2}^{0})$,
$\sqrt{2}Re(S)$] and [$\sqrt{2}Im(H_{1}^{0})$, $\sqrt{2}Im(H_{2}^{0})$, $%
\sqrt{2}Im(S)$]. However, in general case where $h_{i}\neq \left\langle
h_{i}\right\rangle $, such a mixing does exist and it is proportional to $%
\sin \delta $, as shown in Sec. 2.1, and this mixing vanishes at the ground
state. Then the elements of the scalar field-dependent mass-squared matrix
in the basis [$\sqrt{2}Re(H_{1}^{0})$, $\sqrt{2}Re(H_{2}^{0})$, $\sqrt{2}%
Re(S)$, $\sqrt{2}Im(H_{1}^{0})$, $\sqrt{2}Im(H_{2}^{0})$, $\sqrt{2}Im(S)$],
are given by%
\begin{align}
M_{11}^{2}& =\left( g_{1}^{2}+g_{2}^{2}\right) \left(
3h_{1}^{2}-h_{2}^{2}\right) /8+g^{\prime 2}Q_{1}\left(
3Q_{1}h_{1}^{2}+Q_{2}h_{2}^{2}+Q_{s}S^{2}\right) /2+\lambda ^{2}\left(
h_{2}^{2}+S^{2}\right) /2+m_{1}^{2},  \notag \\
M_{12}^{2}& =\left( \lambda ^{2}-\left( g_{1}^{2}+g_{2}^{2}\right)
/4+g^{\prime 2}Q_{1}Q_{2}\right) h_{1}h_{2}+Re(\mathcal{Q})S,  \notag \\
M_{13}^{2}& =\left( \lambda ^{2}+g^{\prime 2}Q_{1}Q_{s}\right) h_{1}S+Re(%
\mathcal{Q})h_{2},~M_{14}^{2}=0,~M_{15}^{2}=Im(\mathcal{Q})S,~M_{16}^{2}=Im(%
\mathcal{Q})h_{2},  \notag \\
M_{22}^{2}& =\left( g_{1}^{2}+g_{2}^{2}\right) \left(
3h_{2}^{2}-h_{1}^{2}\right) /8+g^{\prime 2}Q_{2}\left(
3Q_{2}h_{2}^{2}+Q_{1}h_{1}^{2}+Q_{s}S^{2}\right) /2+\lambda ^{2}\left(
h_{1}^{2}+S^{2}\right) /2+m_{2}^{2},  \notag \\
M_{23}^{2}& =\left( \lambda ^{2}+g^{\prime 2}Q_{2}Q_{s}\right) h_{2}S+Re(%
\mathcal{Q})h_{1},~M_{24}^{2}=Im(\mathcal{Q})S,~M_{25}^{2}=0,~M_{26}^{2}=Im(%
\mathcal{Q})h_{1},  \notag \\
M_{33}^{2}& =g^{\prime 2}Q_{s}\left(
3Q_{s}S^{2}+Q_{1}h_{1}^{2}+Q_{2}h_{2}^{2}\right) /2+\lambda ^{2}\left(
h_{1}^{2}+h_{2}^{2}\right) /2+m_{S}^{2},~M_{34}^{2}=Im(\mathcal{Q})h_{2},
\notag \\
M_{35}^{2}& =Im(\mathcal{Q})h_{1},~M_{36}^{2}=0,  \notag \\
M_{44}^{2}& =\left( g_{1}^{2}+g_{2}^{2}\right) \left(
h_{1}^{2}-h_{2}^{2}\right) /8+g^{\prime 2}Q_{1}\left(
Q_{1}h_{1}^{2}+Q_{2}h_{2}^{2}+Q_{s}S^{2}\right) /2+\lambda ^{2}\left(
h_{2}^{2}+S^{2}\right) /2+m_{1}^{2},  \notag \\
M_{45}^{2}& =-Re(\mathcal{Q})S,~M_{46}^{2}=-Re(\mathcal{Q})h_{2},  \notag \\
M_{55}^{2}& =\left( g_{1}^{2}+g_{2}^{2}\right) \left(
h_{2}^{2}-h_{1}^{2}\right) /8+g^{\prime 2}Q_{2}\left(
Q_{2}h_{2}^{2}+Q_{1}h_{1}^{2}+Q_{s}S^{2}\right) /2+\lambda ^{2}\left(
h_{1}^{2}+S^{2}\right) /2+m_{2}^{2},  \notag \\
M_{56}^{2}& =-Re(\mathcal{Q})h_{1},~M_{66}^{2}=g^{\prime 2}Q_{s}\left(
Q_{s}S^{2}+Q_{1}h_{1}^{2}+Q_{2}h_{2}^{2}\right) /2+\lambda ^{2}\left(
h_{1}^{2}+h_{2}^{2}\right) /2+m_{S}^{2},  \label{Mh}
\end{align}%
with $\mathcal{Q}=A_{\lambda }e^{i(\theta -\theta _{0})}/\sqrt{2}$. The
masses of the charged scalars are given in the basis ($H_{1}^{+}$, $H_{2}^{+}
$)%
\begin{align}
M_{11}^{2}& =\left( \left( g_{1}^{2}+g_{2}^{2}\right) h_{1}^{2}-\left(
g_{1}^{2}-g_{2}^{2}\right) h_{2}^{2}\right) /8+g^{\prime 2}Q_{1}\left[
Q_{1}h_{1}^{2}+Q_{2}h_{2}^{2}+Q_{s}S^{2}\right] /2+\lambda
^{2}S^{2}/2+m_{1}^{2},  \notag \\
M_{12}^{2}& =(g_{2}^{2}-2\lambda ^{2})h_{1}h_{2}/4-\mathcal{Q}^{\ast
},~M_{21}^{2}=\left( M_{12}^{2}\right) ^{\ast },  \notag \\
M_{22}^{2}& =\left( \left( g_{1}^{2}+g_{2}^{2}\right) h_{2}^{2}-\left(
g_{1}^{2}-g_{2}^{2}\right) h_{1}^{2}\right) /8+g^{\prime 2}Q_{2}\left[
Q_{1}h_{1}^{2}+Q_{2}h_{2}^{2}+Q_{s}S^{2}\right] /2+\lambda
^{2}S^{2}/2+m_{2}^{2}.  \label{Mc}
\end{align}

\textbf{Charginos and Neutralinos. }The chargino masses are the same as in
the MSSM with replacing $\mu $\ by its effective value $\mu _{eff}=\lambda S$%
. The two chargino $\tilde{\chi}_{1,2}^{\pm }$ masses are given by the MSSM
formula

\begin{align}
m_{\tilde{\chi}_{1,2}^{\pm}}^{2} & =\frac{1}{2}\left[
\lambda^{2}S^{2}/2+M_{2}^{2}+g_{2}^{2}\left( h_{1}^{2}+h_{2}^{2}\right)
\mp\left\{ \left( \lambda^{2}S^{2}/2-M_{2}^{2}+g_{2}^{2}\left(
h_{1}^{2}-h_{2}^{2}\right) \right) ^{2}\right. \right.  \notag \\
& \left. \left. +2g_{2}^{2}\left(
\lambda^{2}h_{2}^{2}S^{2}/2+M_{2}^{2}h_{1}^{2}\right) +2\sqrt{2}%
g_{2}^{2}\lambda M_{2}h_{1}h_{2}S\cos \theta\right\} ^{\frac{1}{2}}\right] ,
\label{cha}
\end{align}
where $M_{2}$ is the $SU(2)$ gaugino mass.

In the neutralino sector, there is an extra $U(1)^{\prime}$ zino and the
higgsino ${\tilde{S}}$ as well as the four MSSM neutralinos. The $6\times6$
mass matrix reads, in the basis ($\tilde{B}^{\prime}$, $\tilde{B}$, $\tilde{%
W_{3}}$, $\tilde{H}_{1}^{0}$, $\tilde{H}_{2}^{0}$, $\tilde{S}$)
\begin{equation}
M_{\tilde{\chi}^{0}}=\left(
\begin{array}{cccccc}
M_{1}^{\prime} & 0 & 0 & g^{\prime}Q_{1}h_{1} & g^{\prime}Q_{2}h_{2} &
g^{\prime}Q_{S}Se^{i\theta} \\
0 & M_{1} & 0 & -{\frac{1}{2}}g_{1}h_{1} & {\frac{1}{2}}g_{1}h_{2} & 0%
\vspace{0.1cm} \\
0 & 0 & M_{2} & {\frac{1}{2}}g_{2}h_{1} & -{\frac{1}{2}}g_{2}h_{2} & 0%
\vspace{0.1cm} \\
g^{\prime}Q_{1}h_{1} & -{\frac{1}{2}}g_{1}h_{1} & {\frac{1}{2}}g_{2}h_{1} & 0
& -\frac{\lambda}{\sqrt{2}}Se^{i\theta} & -\frac{\lambda}{\sqrt{2}}h_{2} \\
g^{\prime}Q_{2}h_{2} & {\frac{1}{2}}g_{1}h_{2} & -{\frac{1}{2}}g_{2}h_{2} & -%
\frac{\lambda}{\sqrt{2}}Se^{-i\theta} & 0 & -\frac{\lambda}{\sqrt{2}}h_{1}%
\vspace{0.1cm} \\
g^{\prime}Q_{S}Se^{-i\theta} & 0 & 0 & -\frac{\lambda}{\sqrt{2}}h_{2} & -%
\frac{\lambda}{\sqrt{2}}h_{1} & 0%
\end{array}
\right) ,  \label{xi}
\end{equation}
where $M_{1}$ and $M_{1}^{\prime}$ are the gaugino masses associated with $%
U(1)$ and $U(1)^{\prime}$, respectively.

\section{Thermal corrections to the bosonic masses}

\begin{figure}[h]
\begin{center}
\includegraphics[width=13.5cm,height=2cm]{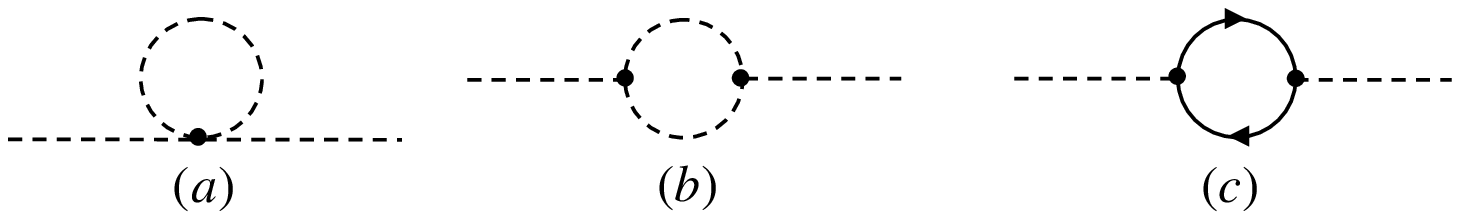}
\end{center}
\caption{\textit{The one-loop diagrams that contribute to the scalar thermal
corrections.}}
\label{Thg}
\end{figure}

All the bosonic fields acquire thermal mass corrections from the three
typical diagrams in (\ref{Thg}-a)-(\ref{Thg}-c). In Fig. \ref{Thg}-c, we
show the fermionic contributions to the scalar mass-squared matrix elements,
while scalar contributions could be evaluated from diagrams in Fig. (\ref%
{Thg}-a) and (\ref{Thg}-b). The gauge contributions could be deduced from
Fig. (\ref{Thg}-a) and (\ref{Thg}-b) just by replacing the internal scalar
legs by gauge ones. The thermal corrections to gauge field masses could be
obtained similarly to the scalar ones by replacing the scalar external legs
by the gauge ones.

When comparing the leading terms from each diagram, $T^{2}/12$, $-mT/8\pi $
and $-T^{2}/24$ respectively, one easily finds that the integral
contribution of (\ref{Thg}-b) is less important with respect to the others,
and therefore one could ignore it when taking the high temperature
expansion. The high temperature expansion is assumed to be a good
approximation for temperatures $m/T<2.2$\ for bosonic and $m/T<1.6$\ for
fermionic masses with an error less than 5\%.

In our model, we are interested in temperatures around the EW scale where
the singlet has already developed its vev, then all the masses and vertices
should be evaluated under the condition%
\begin{equation}
(h_{1},h_{2},S)=(0,0,\upsilon _{x}e^{i\theta _{0}}).  \label{Test}
\end{equation}%
Under this condition, many fields will decouple and therefore do not
contribute to the thermal corrections like $\sqrt{2}Re(S)$, $\sqrt{2}Im(S)$,
$Z^{\prime }$, the stops, the neutralinos and the charginos, and our case
meets exactly the MSSM case \cite{T0}. The internal lines inside the loops
should be mass eigenstates, then the vertices required to evaluate (\ref{Thg}%
) need to be modified by the induced mixing between the interaction states
and the mass eigenstates under the condition (\ref{Test}). In many models,
scalar contributions to the effective potential are generally neglected due
their less relevance with respect to the gauge contributions, therefore, we
will not consider their thermal corrections here. In what follows, we give
the thermal corrections to the scalar, gauge bosons, and squarks.

\textbf{Gauge bosons. }The thermal correction to the charged gauge bosons $%
W^{\pm }$, is given by%
\begin{equation}
\frac{11}{4}g_{2}^{2}T^{2},
\end{equation}%
while the corrections to the mass-squared matrix elements in the basis \{$%
\gamma _{\mu },Z_{\mu },B_{\mu }^{\prime }$\} are given by%
\begin{align}
M_{11}^{2}& =\frac{T^{2}}{6}%
g_{2}^{2}(17g_{1}^{2}+15g_{2}^{2})/(g_{1}^{2}+g_{2}^{2})T^{2},  \notag \\
M_{12}^{2}& =\frac{T^{2}}{6}g_{1}^{2}g_{2}/\sqrt{g_{1}^{2}+g_{2}^{2}}+\frac{%
T^{2}}{6}g_{1}g_{2}(g_{2}^{2}-g_{1}^{2})/(g_{1}^{2}+g_{2}^{2}),  \notag \\
M_{13}^{2}& =-\frac{T^{2}}{12}g_{1}g_{2}g^{\prime }(2Q_{1}-2Q_{2}-3Q_{t})/%
\sqrt{g_{1}^{2}+g_{2}^{2}},  \notag \\
M_{22}^{2}& =\frac{T^{2}}{6}g_{1}g_{2}^{2}/\sqrt{g_{1}^{2}+g_{2}^{2}}+\frac{%
T^{2}}{36}(11g_{1}^{4}+8g_{2}^{2}g_{1}^{2}+9g_{2}^{4})/(g_{1}^{2}+g_{2}^{2}),
\notag \\
M_{23}^{2}& =\frac{T^{2}}{12}Q_{t}g_{1}g^{\prime }+\frac{T^{2}}{12}g^{\prime
}(\left( Q_{1}-Q_{2}\right) \left( 3g_{1}^{2}+g_{2}^{2}\right)
+3Q_{t}g_{2}^{2})/\sqrt{g_{1}^{2}+g_{2}^{2}},  \notag \\
M_{33}^{2}& =\frac{T^{2}}{2}g^{\prime 2}(Q_{1}^{2}+Q_{2}^{2}+Q_{t}Q_{T}),
\label{Thga}
\end{align}%
where $Q_{t,T}$\ is the right-, left-handed top charge under $U(1)^{\prime }$%
. One can obtain the thermal corrections for the MSMM by putting $g^{\prime
}=0$ in (\ref{Thga}), however the result is slightly different than in \cite%
{T0}, because here we consider the scalar contributions.

\textbf{Squarks. }The left- and right-handed squarks are not mixed under the
condition (\ref{Test}), and their thermal corrections to the diagonal
elements are%
\begin{align}
m_{\tilde{t}_{L}}^{2}& =\frac{4}{9}g_{s}^{2}T^{2}+\frac{1}{4}g_{2}^{2}T^{2}+%
\frac{1}{108}g_{1}^{2}T^{2}+\frac{1}{6}y_{t}^{2}T^{2},  \notag \\
m_{\tilde{t}_{R}}^{2}& =\frac{4}{9}g_{s}^{2}T^{2}+\frac{4}{27}g_{1}^{2}T^{2}+%
\frac{1}{3}y_{t}^{2}T^{2},  \label{Thsq}
\end{align}%
where $g_{s}$ and $y_{t}$\ are the strong and Yukawa couplings.

\end{document}